# Access to Library Information Resources by University Students during COVID-19 Pandemic in Africa: A Systematic Literature Review

*Joyce Charles Shikali[1]*
Directorate of Library Services
Institute of Finance Management
Dar es Salaam, Tanzania
jckanyika@gmail.com

*Paul S. Muneja[2]*
Information Studies Programme
University of Dar es Salaam
pmuneja@udsm.ac.tz

*Mohamed. Kassim [3]*
Information Studies Programme
University of Dar es Salaam
mohdie2@yahoo.com

**Abstract**

*The study examined access to library information resources by university students during the outbreak of the COVID-19 pandemic in 2020. Specifically, the study sought to identify the measures adopted by academic libraries to ensure the smooth delivery of library information resources to patrons, particularly students, identify technological tools that were employed by libraries to facilitate access to library information resources. Not only that but also, the study investigated the challenges faced by students in accessing library information resources. A systematic literature review approach following PRISMA guidelines was employed to investigate the findings of the relevant literature on the subject. The keyword search strategy was employed to search for relevant literature from four scholarly databases Scopus, Emerald, Research4life and Google Scholar. The relevant 23 studies were included fulfilling the set inclusion criteria. The presentation of the findings was arranged in a tabular form to provide a summary of each article to facilitate easy analysis and synthesis of results. The findings of this study revealed that the majority of the reviewed studies indicate that, during the COVID-19 pandemic many academic libraries in Africa adopted different approaches to facilitate access to library information resources by university students including expanding access to electronic resources off-campus, virtual reference services, circulation and lending services. To support access to different*

*library services and information resources academic libraries in Africa used various digital technological tools like social media, library websites, email and video conferencing. Moreover, the study revealed that limited access to internet services and ICT devices, inadequate electronic library collection and inadequate digital and information literacy were the major challenges faced by many university students in accessing library resources during the pandemic. This study recommends investment in ICT infrastructures and expanding electronic resource collections which are vital resources in the digital era.*

**Keywords:** Access, library information resources, academic library, university students, COVID-19 pandemic, Africa

**Introduction**
Academic libraries are considered as the center of knowledge and innovation as they engage in collecting and disseminating knowledge both print and electronic resources (Martzouko, 2020). However, abrupt socio-economic changes negatively affect library service provision as well as the patrons. The world is currently recovering from a very serious COVID 19 pandemic that has had impact on all walks of life including libraries. The emergency of COVID 19 has had a serious impact on libraries as it forced them to close (Tammaro, 2020). The outbreak of COVID-19 posed challenges to academics, researchers and students in Universities following their closure to protect against further spread of the virus. Despite the disruption brought by COVID-19 that forced learners to adopt new learning environment, patrons expected their libraries to provide services to meet their information needs (Okonoko *et al.*, 2020; Tsekea & Chigwada, 2020). Patrons expect the library to offer information services through digital communication technologies (Okonoko *et al.*, 2020). For this case, even during the pandemic some libraries could provide access to digital content without distantly.

**Literature Review**
Information and Communication Technology (ICT) has brought the possibility of operating a library beyond the four walls of the physical library. In view of this, academic libraries with well-established technological infrastructure are capable of operating digital library services by providing students with access to digital content regardless of the prevailing challenges. However, academic libraries with unreliable technological infrastructure were completely forced to close their services during the COVID-19 pandemic (Chisita & Chizoma 2021; Dadhe & Dubey, 2020). The library with unreliable technologies will fail to provide information access to its patrons (Chisita & Chizoma 2021; Ali & Gatiti 2020; Ifijeh & Yusuf 2020; Rafiq *et al.*, 2021).

Literature has revealed the way transition from traditional to online library information delivery of has affected academic libraries, especially in the low and middle-income countries including those in African continent (Ali & Gatiti, 2020; Chisita et al., 2022; Fase, Adekoya & Iwari, 2020; Tsekea & Chigwada, 2020). In many African countries, the state of information and communication infrastructures is not well established (Tsekea & Chigwada, 2020). In addition, academic libraries in Africa have been facing several social, economic and technological challenges prior to the COVID-19 pandemic as a result limiting the utilisation of digital technologies to enhance access to library information resources and services (Ashiq *et al.*, 2022). The current COVID-19 pandemic serves as a wake-up call to academic libraries in Africa to assess the way they can continue providing services to users including patrons during the time when physical libraries are inaccessible. Therefore, information on access to library resources provides useful insight into how academic librarians can restructure their services to support access to library information resources by university students during the closure of physical library services. On the other hand, this information will influence future research and policy makers' decisions to support access to library information resources by university students during the pandemic or any other future emergency.

**Purpose of the Study**
The study analysed the literature to ascertain the kind of information services that were provided by University libraries to their students during the outbreak of the COVID-19 pandemic in Africa. Specifically, the study focused on what types of services were adopted by academic libraries to facilitate the accessibility of library information resources by university students following the sudden closure of university campuses because of the COVID-19 pandemic. The study also set out to determine the digital technological tools used by academic libraries to facilitate the accessibility of library information resources and the challenges encountered by university students amid the global pandemic.

**Significance of the Study**
The findings of this study are anticipated to provide insights into academic libraries in Africa as regards to how to offer library services in a time of pandemic or emergency where students cannot pay physical visits to the library buildings as well as the tools and technologies to be harnessed in facilitating student's access to library information resources and the challenge to overcome.

**Research Questions**

This study sought to answer the following research questions:
i.  What types of library services are being offered by academic libraries to students during the outbreak of the COVID-19 pandemic?
ii. What are the digital technological tools being used by academic libraries to facilitate students' access to library information resources during the COVID-19 pandemic?
iii. What are the challenges being faced by university students in accessing library information resources during COVID-19 pandemic?

**Research Methodology**

This study employed a systematic literature review methodology to examine access to library resources by university students during the COVID-19 pandemic. A systematic literature review involves a systematic, transparent and reproducible synthesis of research findings obtained from different empirical findings on a given topic (Davis *et al.*, 2014). The reason for the choice of systematic literature review methodology is to provide a baseline information based on the accumulation of findings from a range of empirical studies which contribute to knowledge development and theory on a given topic (Snyder, 2019). It also uncovers new areas in which further research is needed (Transfield *et al.*, 2003). According to Ayeni *et al.* (2021) systematic literature review helps to combine previous studies that discussed and researched a particular topic and met certain inclusion criteria. Based on this fact the use of a systematic literature review in the current study provides an overview of information on how university students in Africa accessed library information resources during the COVID-19 pandemic as well as contributes to further research on the topic. Various studies were searched, retrieved and analysed systematically to ascertain the kind of services that were offered by libraries during the pandemic as well as the challenges they faced and the measures taken to remedy the situation.

*Search strategy*

The researchers conducted a thorough literature search on four databases (Scopus, Emerald, Research4life and Google Scholar) to identify relevant studies. The literature search was first conducted on August 16, 2022 and updated on December 7, 2022, to account for the latest articles. The custom range of 2020 through 2022 and sort by relevance were the key parameters during the literature search. This custom range was used to cover current research related to the era of the COVID-19 pandemic. The researchers used five main keywords "university students", "academic library", "library information resources", "COVID-19 pandemic" and "Africa" to construct search queries to identify the relevant studies covering the objectives of the study. During the search, the researchers used the Boolean operator "AND" to narrow down research results and the Boolean operator "OR" for expanding the search query.

The search process in each database were as follows:

i. Researchers run a search query in the Scopus database using TITLE-ABS-KEY and then applied the following filters (year of publication 2020-2022; source type-journal; document type-article; language-English). This search resulted in 89 results.
ii. In the Emerald database, researchers just put the search query in a search box then applied the following limiter (search by relevancy, year of publication 2020-2022; source type-journal; document type-article; language-English). This search resulted in 652 results
iii. In Research4life, researchers put the search query in the search box and then refined the search by (The publication year 2020-2022; Scholarly peer-reviewed journal article; field of study-library and information science and language –English). The search resulted in 8 results.
iv. In Google scholar, researchers put the search query in the box and then applied the following filter (Custom range 2020-2022; sort by relevance, pear reviewed article; language- English. This search resulted in 784 results. The Overall results from all four databases were 1,533 records

*Inclusion and exclusion criteria*

The focus of the study was on covering access to library information resources by university students in the COVID-19 pandemic era in Africa. The inclusion criteria were:

i. Only articles published in peer review journals. The peer-reviewed journal articles were considered as the ones that maintain standards and enhance the quality of the work because they undergo a quality check mechanism which helps to strengthen the credibility of scholarly publications.
ii. Research articles published from 2020 to 2022 on access to library information resources by university students in Africa. Taking into consideration that COVID-19 began in December 2019, therefore, it is possible that immediately after 2019 research on this subject began.

iii. Research article published in the English language because English is the major research language in the field of Library and Information Science.
iv. Studies covering more than one aspect of access to library information resources by university students during the COVID-19 pandemic in Africa.

The exclusion criteria were based on studies not published in peer review journals, not published in the English language and not discussed access to library information by university students in the COVID-19 pandemic era in Africa as well as studies covering only one aspect of this study.

*Selection of studies*

A systematic review of relevant literature was conducted following the Preferred Reporting Items for the Systematic Review and Meta-analysis (PRISMA) by (Moher *et al.*, 2009). The PRISMA aims at helping authors to improve the reporting of systematic reviews and meta-analyses. It provides a checklist that guides the researchers in the identification of relevant material, screening, eligibility and included studies for literature review synthesis. The PRISMA assumes that the quality of the systematic review and meta-analysis depends on the scope and the quality of the included studies (Moher *et al.*, 2009). The PRISMA protocol for systematic review and meta-analysis has four stages which include identification, screening, eligibility and inclusion. In the first stage, a total of 1533 articles were identified from the four databases. In the second stage, the title and abstract of articles were screened by using the first stage of inclusion and exclusion criteria. The potential articles which fit in the study were identified and saved into Mendeley library where duplicates were removed. In the third stage, the full text of identified articles was screened for eligibility by using the second stage of eligibility criteria. A critical evaluation of the full text of the selected articles was performed to check whether they meet the objective of the study. The final stage presents 23 articles that were included for systematic review and analysis.

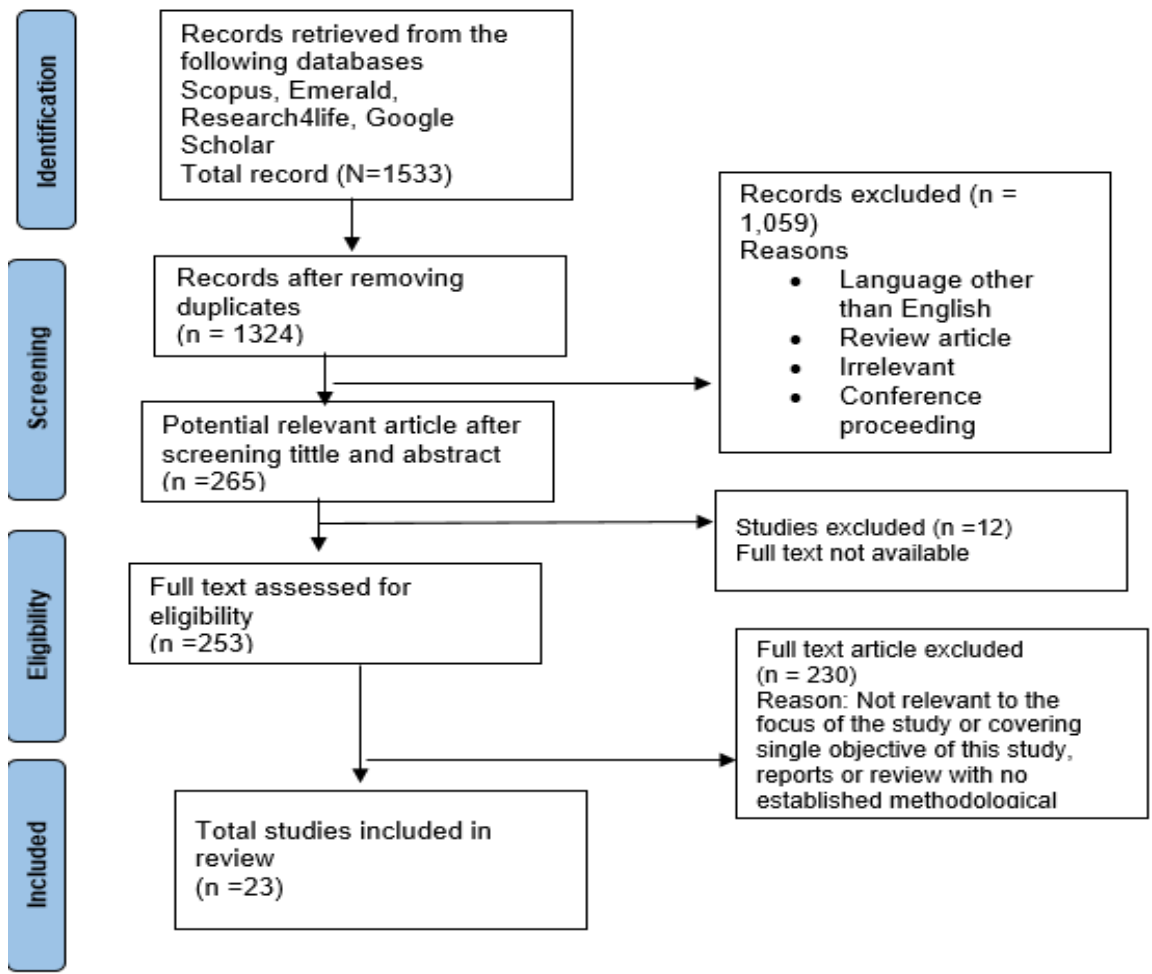

**Figure 1**: Four phase flow diagram of selection procedure of included studies

**Data extraction:** A tabular approach was used to provide a summary for each eligible study. Data extraction (See Table 1) which includes authors' names, the methods used, types of eservices offered by academic libraries during COVID-19, digital technological tools implemented and the challenges faced by students in accessing library resources was used to extract data summary from 23 reviewed articles.

**Results and Interpretation**

This section presents the summary of data extracted from 23 reviewed articles based on the specific objectives of the study. The section includes the synthesis of themes identified based on each specific objective and research question.

**Objective 1: Types of services offered by academic libraries to support students' access to library information resources during COVID-19 pandemic**

Table 2 presents different types of services which were offered by academic libraries in Africa to facilitate students' access to library resources during the closure of higher learning institutions following the outbreak of COVID-19 pandemic.

**Table 2: Types of services provided by academic library during COVID-19 pandemic**

| Types of service provided | Frequency | Percent |
|---|---|---|
| Remote access to library resources | 20 | 87 |
| Virtual reference services | 18 | 78 |
| Circulation and lending services | 8 | 35 |
| Online user education programs | 8 | 35 |
| Research support services | 5 | 22 |

These results show that the majority (87%) of academic library services were offered through remote access to library resources. In other words, remote access was preferred possibly due to the closure of universities and some precautions related to lockdown and social distancing. Apart from that, a significant percentage (78%) of academic libraries applied virtual reference services in offering library services to users. Other services offered by academic libraries during the pandemic include circulation and lending services, online user education programme and research support services.

**Objective 2: Digital technological tools used by academic libraries to facilitate access to library information resources by university students during the outbreak of COVID-19**

During COVID-9, academic libraries resorted to the application of different digital technology tools to optimise remote access to library information resources and services at the same time maintain social distance. There were numerous tools used by academic libraries in supporting university students' access to library information resources during the COVID-19 lockdown. Table 3 presents the findings on the digital technological tool used by academic libraries in supporting library information resources during COVID-19.

**Table 3: Digital technological tools used by academic libraries to facilitate access to library information resources during COVID-19 Pandemic**

| Digital tools used to facilitate access during COVID-19 pandemic | Frequency | Percentage |
|---|---|---|
| Social media | 14 | 61 |
| Library website | 13 | 57 |
| Ask a librarian and live chat | 8 | 35 |
| Email | 7 | 30 |
| Library OPAC | 5 | 22 |
| Remotex and EZProxy | 3 | 13 |
| Video conferencing | 3 | 13 |

These results show that majority (61%) of academic libraries used social media to support university students' access to library information resources. As such, application of social media mostly used by academic libraries as it allows a mode of communication without being physically attached. Besides, a significant percentage (57%) of academic libraries prefer usage of library websites in supporting university students' access to library information resources. Apart from that, other electronic tools that do not necessarily require physical presence in the library in accessing library information resources were applied such as Ask a librarian and live chat, email, library OPAC, Remotex and EZProxy and video conferencing.

## Objective 3: Challenges faced in accessing library information resources during COVID-19 pandemic

Providing services to university students during COVID-19 faced myriad challenges. Table 4 presents various challenges faced by university students in accessing library information resources during COVID-19.

**Table 4: Challenges faced by university students in accessing library information resources during COVID-19**

| Challenges | Frequency | Percentage |
|---|---|---|
| Limited access to internet services | 19 | 82.6 |
| Inadequate library e-resources collection | 10 | 43.5 |
| Inadequate digital and information literacy skills among students | 9 | 39 |
| Limited off-campus access to library e-resources | 5 | 22 |
| Lack of knowledge on library resources which can be accessed remotely | 4 | 17 |
| Limited interaction between librarians and students | 4 | 17 |

These results show that the majority (82.6%) of reviewed studies indicate that limited access to internet service is the major challenge faced by students in accessing remote library resources during COVID-19. Apart from that, ten (10) out of twenty-three (23) reviews indicated that inadequate library e-resources collection was another challenge faced by university students in accessing library resources during the pandemic. This has been influenced by inadequate funds among academic libraries in Africa to subscribe to more e-resources to meet the increasing students' demand for e-resources. Also, inadequate digital and information literacy skills among students and limited off-campus access to library electronic resources hindered students' remote access to library resources during the COVID-19 pandemic. Lack of knowledge on library resources that can be accessed remotely and limited interaction between librarians and patrons due to social distance and lockdown each mentioned by four out of 23 reviewed studies.

**Discussion**

The study sought to examine access to library information resources by university students during the outbreak of the COVID-19 pandemic in Africa. Specifically, the study identified types of services offered by academic libraries to support university students' access to library information resources, identified digital technological tools employed by academic libraries and the challenges faced in accessing library information resources during the COVID-19 pandemic.

The findings of the study revealed that during an emergency that requires temporary closure of university campuses adjustment of physical library services to virtual library services and other innovative services is inevitable for academic libraries in Africa to facilitate off-campus electronic access to library resources to ensure students' continuity of access to library information resources. On this, remote access to library information resources such as newspapers, e-books, e-journals, past examination papers, institution repositories, online public catalogs, streaming media, research guides online databases, library electronic resources and other educational contents through various digital technological tools are pivotal during an emergency time like the case of COVID-19 (Chisita *et al.*, 2022; Tsekea & Chigwada, 2020; Mbambo-Thata, 2020). On the other hand, provision of virtual library services by academic libraries during emergency like COVID-19 pandemic is virtual. There has been a significant increase in the use of online reference services to sustain interaction with patrons and answer patrons' queries, to support remote access to library information resources (Mathabela, 2021;

Tsekea & Chigwada, 2020), to create awareness and promote library resources and services in a digital environment (Chisita & Chizoma, 2020; Chisita et al., 2022; Ifijeh & Yusuf, 2020; Abubakar, 2021).

Some libraries introduced online interlibrary loans and document delivery to support remote access to library resources, especially for students who were not able to access online information resources (Ifijeh & Yusuf, 2020; Chigwada, 2022; Magut, 2022). Other libraries extend the book loan period and waive fines to allow students to stay with books and other information resources during the lockdown period and avoid the accumulation of fines during the lockdown period (Mathabela, 2021). In exceptional ways, Curbside book pick up was another new service offered by some academic libraries in South Africa to facilitate access to print library resources for students who were not able to utilize digital facilities to access library resources during the lockdown (Mashiyane & Molepo, 2021). To return borrowed library information resources, some libraries used book drop boxes whereby students can return borrowed books and other library materials without physical contact with librarians (Tsekea & Chigwad, 2020; Chisita & Chizoma, 2020; Chisita *et al.*, 2022).

On the part of the tool used to facilitate access to library information resources during the COVID-19 pandemic, the outbreak of the COVID-19 pandemic triggered a keen interest among academic libraries in Africa to adapt digital technology tools to reach users during the closure of physical library services. Digital technological tools have been used to allow students to access electronic library resources anywhere at any time following the temporal closure of library buildings. In this regard, Chisita *et al.* (2022) pointed out that the closure of libraries and lockdown create a physical barrier between librarians and patrons resulting in increased demand for digital technology tools to facilitate access to library services and resources to avoid total closure and suspension of services. These tools include social media, library website and email, ask a librarian and live chat, videoconferencing and other software like Remotex and EZProxy. The utilisation digital technology tools also helped to bridge the distance between library staff and patrons during lockdown. The findings of this study revealed that many academic libraries in Africa used social media tools like (Facebook, Twitter, WhatsApp, YouTube, Mayspace, Telegram and Blogs) to offer different library services and provide link to library e-resources. Virtual reference services were provided through several digital tools like social media, Lib-guide, chat facilities on website, phone number and email (Abubakar 2021; Ifijeh, 2020; Mbambo-Thata, 2021). For instance, the University of Lesotho library provided reference services through chat facilities while subject specific queries were answered by subject librarians (Mbambo-Thata, 2021). On the other hand, different web conferencing tools such as Zoom, WebEx, BigBlueButton, Google meet, Microsoft team, instructional video guides on the library websites and social media have been used by academic libraries to facilitate delivery of online user education programs during lockdown Tsekea & Chigwada, 2020, Chisita & Chizoma, 2020; Chigwada 2022). To support this, Abubakar (2021) asserts that social media implementations and daily usage increasingly become among librarians around the global during the COVID -19 pandemic. On the other hand, Ifijeh and Yusuf (2020) emphasise that university libraries could leverage the use of social media to promote reference services during the COVID-19 pandemic.

Despite the initiatives made by academic libraries to provide support on access to library resources during the closure of the library following the outbreak of COVID-19, students in Africa faced a number of challenges that limited smooth access to information resources offered by academic libraries through various digital platforms. Studies have indicated that limited access to internet service and ICT devices is the major challenge that hindered students' access to remote online information resources during the COVID-19 pandemic. It was established that some university students in many African countries had challenges in having access to reliable internet services and computer facilities to support online access to library information resources during the lockdown. Nwosu (2021) asserts that access to the internet and other digital facilities for many university students in African countries is made available on their host university campuses, but limited access to physical library buildings presents a challenge for students to purchase their own gadgets such as laptops and computers to access the online library information resources and other digital services. Due to financial constraints, many students were not able to meet the cost of purchasing computer or laptop and meet data cost (Mbambo-Thata, 2020; Martizirofa *et al.* 2021). Furthermore, the findings of the study identified that inadequate library e-resources and limited off-campus access to electronic library information resources hindered students' access to library information resources during the pandemic period. As such, not all academic libraries in Africa managed to offer remote access to library resources during the COVID-19 pandemic due to continuous budget cuts and poor technological infrastructure to support off-campus access (Tsekea & Chigwada, 2020). Inadequate funding further deters technological tools for accessing library information resources and the library collection. High electronic resources subscription costs resulted in the limited library collection, especially e-resources. Mathabela (2020) discloses that one of the challenges which contributed to the limited digital collection at the University of Eswatini was the lack of finance for e-resources subscriptions. Along with that, inadequate digital and information literacy skills among university students, lack of knowledge of the available library resources which can be accessed remotely and limited interaction with librarians contributed to difficulties encountered by university students in accessing library information resources during the COVID-19 pandemic.

**Implications of the Study**

*Implication for practice*
This systematic literature review provides insight into access to library information resources by university students during the COVID-19 pandemic in Africa. The study has revealed that like in other parts of the world, academic libraries in Africa have taken initiatives to facilitate remote access to library information resources and services through digital platforms to ensure that university students are not denied access to scholarly information resources needed to support their learning in a changing digital learning environment following the closure of university campuses due to COVID-19 pandemic. Despite the initiatives taken by academic libraries to support student's access to library information resources, actions need to be taken by library management and parent institutions management in Africa to improve ICT infrastructures, expand library electronic resource collections and impart digital literacy skills to both library professionals and students. Otherwise, access to library information resources will still be problematic in many academic libraries in Africa during the emergency even in the future.

*Implications for policy*
This study provides useful insights to organisational policymakers and academic library directors in Africa in the development of emergency and disaster preparedness policy which will guide academic libraries on how to provide access to library information resources and services to university students during the COVID-19 pandemic and other emergency or disaster in the future.

**Limitations and Recommendations for Further Studies**
The limitation of this study included the selection of databases, language, search strategies and quality assessment of the selected studies. Four scholarly databases (Scopus, Emerald, Research4life and Google Scholar) were selected to extract data for this study. In addition, the gray literature such as conference papers, proceedings, dissertations, reports, discussions, etc was not included in this study. Therefore, it is possible that some potential records and studies published in other databases might be missed. Furthermore, selected keywords were used to construct search queries. It is possible some records did not be included due to missing keywords or limitations of a search query.

This systematic review recommends further studies should be conducted to assess access to library information resources and services by university students in a specific country accounting for geographical locations and language. This will help to provide a good understanding of the current status regarding access to library information resources and services by university students during times of health emergencies like the COVID-19 pandemic. Also, another study should focus on the post-pandemic experience of different categories of libraries regarding the provision of services to users.

**Conclusions and Recommendations**
This systematic literature review analysed access to library information resources by university students during the outbreak of the COVID-19 pandemic in Africa. Specifically, the study sought to identify the measures adopted by academic libraries to ensure the smooth delivery of library information resources to patrons, particularly students, identify technological tools that were employed by libraries to facilitate access to library information resources; also, the study investigated the challenges faced by students in accessing library information resources. It is evident that the outbreak of the COVID-19 pandemic has challenged traditional library services worldwide. To stay relevant academic libraries in Africa have to expand library services through digital technology platforms to ensure that library information resources are accessible to students despite the closure of physical library services. Some innovative services are required to be adopted by libraries in Africa to offer off-campus access to library information resources during the closure of physical library services at the same time promote the use of electronic resources more than it was before. On the other hand, academic libraries should put more emphasis on providing digital and information literacy training to students to impart them with the required skills to be able to navigate through the changing online information landscape brought about by the COVID-19 pandemic. Moreover, academic libraries should be well equipped to deal with emergency situations by investing in ICT infrastructure and expanding electronic resource collection to continue supporting students' access to library information resources during emergency times as in the case of the COVID-19 pandemic. The government also should bridge the digital divide gap by supporting students' access to internet services and other digital facilities for students to be connected and access electronic information resources at home.

**References**

Abubakar, M. K. (2021). Implementation and Use of Virtual Reference Services in Academic Libraries during and post COVID-19 Pandemic: A Necessity for Developing Countries. *Library Philosophy & Practice*.

Ali, M. Y., & Gatiti, P. (2020). The COVID-19 (Coronavirus) pandemic: reflections on the roles of librarians and information professionals. *Health information & libraries journal*, *37*(2), 158-162.

Asimah, A. P. A., Dzogbede, O. E., & Akaba, S. (2021). Digital library usage during the covid-19 pandemic. *Library Philosophy and Practice*, 0_1-15.

Ayeni, P. O., Agbaje, B. O., & Tippler, M. (2021). A systematic review of library services provision in response to COVID-19 pandemic. *Evidence Based Library and Information Practice*, *16*(3), 67-104.

Bhati, P., & Kumar, I. (2020). Role of library professionals in a pandemic situation like COVID-19. *International Journal of Library and Information Studies*, *10*(2), 33-48.

Chisita, C. T., & Chizoma, U. S. (2021). Rethinking academic library space amidst the COVID-19 pandemic in South Africa: preparing for the future. *Information Discovery and Delivery*.

Chisita, C. T., Chiparausha, B., Tsabetse, V., Olugbara, C. T., & Letseka, M. (2022). Remaking academic library services in Zimbabwe in the wake of COVID-19 pandemic. *The Journal of Academic Librarianship*, *48*(3), 102521.

Chigwada, J. (2022). The Impact of COVID-19 on Academic Library Service Delivery in Zimbabwe.

Dadhe, P. P., & Dubey, M. N. (2020). Library Services Provided During COVID-19 Pandemic: Content Analysis of Websites of Premier Technological Institutions of India. *Library Philosophy & Practice*.

Davis, J., Mengersen, K., Bennett, S., & Mazerolle, L. (2014). Viewing systematic reviews and meta-analysis in social research through different lenses. *SpringerPlus*, *3*(1), 1-9.

Fasae, J. K., Adekoya, C. O., & Adegbilero-Iwari, I. (2020). Academic libraries' response to the COVID-19 pandemic in Nigeria. *Library Hi Tech*.

Gilbert, N. (2021) Access to library materials in nacademic libraries by students during COVID-19 era: A case of Open University learning institution.

Ifijeh, G., & Yusuf, F. (2020). Covid–19 pandemic and the future of Nigeria's university system: The quest for libraries' relevance. *The Journal of Academic Librarianship*, *46*(6), 102226.

Kasa, M. G., & Yusuf, A. (2020). Experience of an Academic Library during the covid-19 pandemic. *Library Philosophy and Practice*, *1*.

Kumah, M. A., Ocran, T. K., & Parbie, S. K. (2021). COVID 19 Synchronicity: Evaluating Virtual Reference Services of some Academic Libraries in Africa.

Magut, H., & Kiplagat, S. (2022). *Covid-19 Pandemic and the Role of Kenyan University Libraries in Online Education.*

Matizirofa, L., Soyizwapi, L., Siwela, A., & Khosie, M. (2021). Maintaining student engagement: The digital shift during the coronavirus pandemic a case of the library at the University of Pretoria. *New Review of Academic Librarianship*, *27*(3), 364-379.

Martzoukou, K. (2020). Academic libraries in COVID-19: a renewed mission for digital literacy. *Library management.*

Mathabela, N. N. (2021). Library Services during the Covid-19 Pandemic: A Case of the University of Eswatini (UNESWA). *The Christian Librarian*, *64*(1), 12.

Mashiyane, D., & Molepo, M. (2021). Curbside book pick-up services during a time of crisis in South African University Libraries.

Mbambo-Thata, B. (2020). Responding to COVID-19 in an African university: the case the National University of Lesotho library. *Digital Library Perspectives.*

Moher, D., Liberati, A., Tetzlaff, J., Altman, D. G., & PRISMA Group*. (2009). Preferred reporting items for systematic reviews and meta-analyses: the PRISMA statement. *Annals of internal medicine*, *151*(4), 264-269.

Mnzava, E. E., & Katabalwa, A. S. (2021). Library websites during the COVID-19 pandemic. *Alexandria*, *31*(1-3), 12-22.

Nwosu, J. C., & Asuzu, C. M. Library Services and Information Access in a Time of Pandemic: How are Academic Librarians in Nigeria Carrying out Library Services?

Okike, B. I. (2020). Information dissemination in an era of a pandemic (COVID-19): librarians' role. *Library Hi Tech News.*

Okonoko, V. N., Abba, M. A., & Arinola, A. E. (2020). Users' Expectation of Library Services and Resources in the COVID 19 Pandemic Era: A Comparative Study of Two Academic Libraries in Nigeria. *Library of Progress-Library Science, Information Technology & Computer*, *40*(2).

Ogunbodede, K., Ambrose, S. E., & Idubor, I. (2021). Undergraduate students' use of electronic resources amid the COVD-19 pandemic lockdown. *Library Philosophy and Practice*, *6434*.

Omeluzor, S. U., Nwaomah, A. E., Molokwu, U. E., & Sambo, A. S. (2022). Dissemination of information in the COVID-19 era in university libraries in Nigeria. *IFLA journal*, *48*(1), 126-137.

Rafiq, M., Batool, S. H., Ali, A. F., & Ullah, M. (2021). University libraries response to COVID-19 pandemic: A developing country perspective. *The Journal of Academic Librarianship*, *47*(1), 102280.

Shonhe, L. (2022). Covid-19 and African Libraries: Perspectives through Literature from the Library Philosophy and Practice (E-Journal*). Library Philosophy and Practice (e-journal),* 6940.

Snyder, H. (2019). Literature review as a research methodology: An overview and guidelines. *Journal of business research*, *104*, 333-339.

Thuo, M. W. (2021). Adjusting Academic Library Services to Covid 19 Prevention Protocols.

Tranfield, D., Denyer, D., & Smart, P. (2003). Towards a methodology for developing evidence-informed management knowledge by means of systematic review. *British journal of management*, *14*(3), 207-222.

Tsekea, S., & Chigwada, J. P. (2020). COVID-19: Strategies for positioning the university library in support of e-learning. *Digital Library Perspectives*.